\documentclass[aps,prl,twocolumn,showpacs,showkeys]{revtex4}

\usepackage{graphics}
\usepackage{amssymb}
\usepackage{amsmath, amsthm}

\newcommand{\be}{\begin{equation}}
\newcommand{\ee}{\end{equation}}

\begin{document}

% Use the \preprint command to place your local institutional report
% number in the upper righthand corner of the title page in preprint mode.
% Multiple \preprint commands are allowed.
% Use the 'preprintnumbers' class option to override journal defaults
% to display numbers if necessary
%\preprint{}

%Title of paper
\title{$R^n$ gravity and the chameleon}

% repeat the \author .. \affiliation  etc. as needed
% \email, \thanks, \homepage, \altaffiliation all apply to the current
% author. Explanatory text should go in the []'s, actual e-mail
% address or url should go in the {}'s for \email and \homepage.
% Please use the appropriate macro foreach each type of information

% \affiliation command applies to all authors since the last
% \affiliation command. The \affiliation command should follow the
% other information
% \affiliation can be followed by \email, \homepage, \thanks as well.
\author{Valerio Faraoni}
\email[vfaraoni@ubishops.ca]{}
%\homepage[]{Your web page}
%\thanks{}
%\altaffiliation{}
\affiliation{Physics Department, Bishop's University\\
Sherbrooke, Qu\'ebec, Canada J1M~1Z7
}

%Collaboration name if desired (requires use of superscriptaddress
%option in \documentclass). \noaffiliation is required (may also be
%used with the \author command).
%\collaboration can be followed by \email, \homepage, \thanks as well.
%\collaboration{}
%\noaffiliation

%\date{\today}

\begin{abstract}
The Solar System bounds on $R^n$  gravity are often ignored in 
the literature by invoking the chameleon mechanism. We show 
that in order for the 
latter to work, the exponent $n$ must be ridicolously close to 
unity  and,  therefore, these theories are severely constrained.
\end{abstract}

% insert suggested PACS numbers in braces on next line
%Mathematical and relativistic cosmology
% 04.50.Kd modified theories of gravity
% 04.50.-h=other topics in GR and gravitation
%98.80.Jk Mathematical and relativistic aspects of cosmology
% insert suggested keywords - APS authors don't need to do this

%\maketitle must follow title, authors, abstract, \pacs, and \keywords
\maketitle

%\section{}
% Put \label in argument of \section for cross-referencing
%\section{\label{}}
%\subsection{}
%\subsubsection{}

% If in two-column mode, this environment will change to single-column
% format so that long equations can be displayed. Use
% sparingly.
%\begin{widetext}
% put long equation here
%\end{widetext}

The acceleration of the cosmic expansion discovered with type Ia 
supernovae \cite{supernovae} still lacks a satisfactory 
explanation. The hypothetical dark energy wich is supposed to 
drive this acceleration is an {\em ad hoc} explanation: it 
cannot be detected directly in the laboratory and is  
extremely exotic due to its negative pressure $P$. Its equation 
of 
state should be $P\simeq -\rho$ (where $\rho$ is the 
comoving energy  density) and phantom energy, which opens 
the door to much trouble with its instabilities and 
thermodynamical 
behaviour, is not at all excluded by the observations. Much 
theoretical effort has 
gone into proposing an abundance of models for dark energy and to 
constrain it observationally (see \cite{AmendolaTsujikawabook} 
for a detailed discussion and for references).  An 
alternative approach 
consists of dispensing with dark energy and postulating, instead, 
that Einstein's theory of General Relativity (GR) fails at the 
largest  scales and that, with the cosmic acceleration, we have 
detected departures from the 
expected GR behaviour. This proposal \cite{CCT, CDTT} has led 
to a revival  of $f(R)$ or ``modified'' gravity, described by the 
action
\be\label{f(R)action}
S=\frac{1}{2\kappa} \int d^4x \sqrt{-g} f(R) +S^{(m)} \,,
\ee
where $g$ is the determinant of the spacetime metric $g_{ab}$, 
$R$ is the Ricci scalar, $\kappa=8\pi G$, $G$ is Newton's 
constant, and $S^{(m)}$ is the matter  action.  This class of  
theories, which reduces to GR for a linear  function $f(R)$, 
comes in three versions:  metric, Palatini, and metric-affine 
formalisms (see  \cite{SotiriouFaraoni10, DeFeliceTsujikawa} for 
reviews and  \cite{otherreviews} for 
introductions). The more complicated 
metric-affine formalism \cite{metricaffine} is not fully 
developed yet and has 
seen little use in cosmology. Inside matter the Palatini 
formalism, in which the metric  and the connection are treated 
as independent variables, is  riddled with problems unless its 
field equations  get modified by higher order terms 
\cite{BarausseSotiriouMiller} 
and, therefore, we will discuss here only the metric formalism, 
in which the connection is the metric connection (the distinction 
between metric and Palatini formalisms is irrelevant for GR, but 
the two variations produce inequivalent field 
equations for non-linear $f(R)$ functions). 

Metric $f(R)$ gravity contains a scalar degree of freedom,  
identified with $\phi \equiv f'(R)$. In fact, metric $f(R)$ 
gravity is a Brans-Dicke theory \cite{BransDicke} with parameter 
$\omega=0$ and a special potential for the Brans-Dicke field 
$\phi$ \cite{STequivalence}. Starting from 
the action~(\ref{f(R)action}) and   introducing a new field 
$\chi$, the action 
\be
S=\frac{1}{2\kappa} \int d^4x \sqrt{-g} \left[ f(\chi) 
+f'(\chi)\left( R-\chi\right) \right] +S^{(m)} 
\ee
is dynamically equivalent to~(\ref{f(R)action}). Variation with 
respect to $\chi$ yields $f''(\chi)\left( R-\chi \right)=0$ and 
$ \chi=R $ if $ f''(R)\neq 0$, and the action~(\ref{f(R)action}) 
is 
reproduced. If we define the field $\phi \equiv f'(\chi)$ and set
\be\label{potential}
V(\phi)=\chi(\phi)\phi - f(\chi(\phi)) \,,
\ee
the action becomes \footnote{We follow the notations of 
Ref.~\cite{Wald}.}
\be
S=\frac{1}{2\kappa} \int d^4x \sqrt{-g} \left[ \phi R -V(\phi) 
\right] +S^{(m)} \,,
\ee
an $\omega=0$ Brans-Dicke theory \cite{BransDicke}.  

Many choices for the function $f(R)$ have appeared in the 
literature, and there are viable ones which satisfy 
both theoretical 
viability criteria (such as correct cosmological dynamics, smooth 
transition between different cosmological eras, well-posed 
initial value problem, stability, correct weak-field 
limit and dynamics of  cosmological perturbations) and 
experimental constraints \cite{SotiriouFaraoni10, 
DeFeliceTsujikawa}. There is a large body of  literature 
(\cite{rnliterature1, rnliterature2} and references therein) on 
the choice $f(R)=\alpha  R^n $ (where $\alpha>0$ has the 
dimensions of  a mass squared and $n$ is not 
restricted to be an  integer), on which we focus. Let us be 
clear on the terminology here: often, the literature refers to 
the theory described by $f(R)=R+\alpha R^2$ motivated by quantum 
corrections to the Einstein-Hilbert Lagrangian  as 
``$R^2$-gravity'' (and, consequently, to $f(R)=R+\alpha R^n$ as 
``$R^n$-gravity''). This is not what we mean here: the term 
``$R^n$-gravity'' in this paper refers strictly to 
the choice $f(R)=\alpha R^n$ and our considerations apply only to 
this class of theories (the prospects appear much better for 
$f(R)=R+\alpha R^n$ theories).

$R^n$ gravity, like any $f(R)$ theory, is subject to 
experimental constraints:  while, from the  mathematical physics 
point of view, it is perfectly acceptable to study this theory 
as a toy model in order to obtain  analytical  or qualitative 
insight on exact solutions, or on the 
role that the scalar degree of freedom $f'(R)$ may play in 
modifying GR, or to replace the full theory $f(R)=R+\alpha 
R^n$ with $\alpha R^n$ (which is mathematically 
easier to handle) for $n>0$ in the strong gravity regime, the 
exponent $n$ is not an 
entirely free parameter if the theory is meant to constitute  a 
realistic  alternative to dark energy.

Let us first consider two basic theoretical requirements  
associated with stability. First, avoiding the Dolgov-Kawasaki 
instability \cite{DolgovKawasaki} leads to  $f''(R) \geq 0 $ 
\cite{mystability,  NojiriOdintsovstability} which 
corresponds to $n\leq 0$ or $n\geq 1$ for 
$f(R)=\alpha R^n$. Second, on time scales shorter than the Hubble 
time, one  models the present  universe as a de Sitter one, and 
de Sitter space is usually found to 
be a late-time attractor in $f(R)$ and dark energy 
models. Therefore, it is important that de Sitter space be 
stable, too, which provides the second criterion. 
The mass $m$  of the scalar field $\phi=f'(R)$ in de Sitter space 
is given by
\be \label{masssquared}
m^2= \frac{1}{3} \left( \frac{f_0'}{f_0''}-R_0 \right) \,,
\ee
where a zero subscript denotes quantities evaluated in the de 
Sitter space with Ricci scalar $R_0$. Eq.~(\ref{masssquared}) has 
been  derived in a  variety of ways, including the weak-field 
limit \cite{Mulleretal90, NavarroVanAcoleyen07, Chibaetal},  
gauge-invariant perturbation analyses of de Sitter space 
\cite{FaraoniNadeau05}, and  calculations of the propagator of 
$f(R)$ gravity in a locally flat background 
\cite{NunezSolganik04}. For $f(R)=\alpha R^n$, it  is 
\be
m^2=\frac{(2-n)}{3(n-1)}\, R_0 \,,
\ee
and the requirement that the field $\phi$ be non-tachyonic is 
equivalent to $1\leq n \leq 2$. We take the parameter $n$  
in the intersection of these two  intervals $1\leq 
n \leq 2$ (bounded from below by GR).

In order for $f(R)=\alpha R^n$ to provide a realistic alternative 
to dark energy, it also needs  to satisfy the available  
experimental constraints. Writing $n \equiv 
1+ \delta $, light deflection does not provide bounds 
\cite{CliftonBarrow05a, Lubinietal} but the precession of 
Mercury's perihelion yields the stringent limits
\cite{CliftonBarrow05a, 
CliftonBarrow06,
BarrowClifton06,
Zakharovetal06}
\be \label{limitsondelta}
\delta=\left( 2.7 \pm 4.5\right) \cdot 10^{-19}  \,.
\ee
This constraint is often ignored in studies of $R^n$ gravity 
\cite{rnliterature2}, 
based on the belief that the Solar System limits 
are circumvented because in the weak-field limit of 
general $f(R)$ gravity, the effective 
degree of freedom $\phi=f'(R)$ is endowed with  a range which may 
be very small at Solar system densities and much larger at 
cosmological densities. This feature would  enable effects on 
cosmological scales but would shelter $\phi$ from  
the experimental bounds in the Solar System (the 
chameleon mechanism at work, see below). This argument is 
misleading: let us examine how it applies to the weak 
field limit of $f(R)$ gravity in general, and then discuss the 
specific $R^n$ theory. 

The weak-field limit of $f(R)$ gravity has been studied by 
various authors \cite{Chibaearly, Chibaetal, Olmo07, 
FaraoniTremblaycomment}. Based on the equivalence between 
metric $f(R)$  and $\omega =0$ Brans-Dicke gravity and on the 
Cassini bound $|\omega|>40000$ \cite{Bertottietal03}, early work 
dismissed all $f(R)$ theories as unviable \cite{Chibaearly}. 
However,  the fact was missed that the Cassini limit 
only applies  to a Brans-Dicke field with range larger than, or 
comparable to, the size of Solar System experiments, while the 
effective mass and range of the 
scalar  field  $\phi=f'(R)$ depend on the background curvature 
$R$, hence on the energy  density of the environment. This is 
the chameleon mechanism originally discovered in quintessence 
models of dark energy \cite{Khoury}, and later rediscovered 
in modified gravity \cite{Faulkner}. The 
chameleon  mechanism is not imposed to fine-tune the theory and 
evade the  experimental limits: it is contained naturally in 
$f(R)$ gravity and whether it works or not depends on the 
specific theory considered.

In the weak-field limit of $f(R)$ theories 
\cite{Chibaetal, Olmo07},  one considers a spherically 
symmetric, weakly  gravitating,  perturbation of mass $M$ of a 
cosmological space.  In an  adiabatic approximation, the 
background is taken 
to be a de  Sitter space (with constant curvature, 
$R_{ab}=R_0g_{ab}/4$, and 
$R_0=12H_0^2$), which is a solution of $f(R)$ gravity subject to 
the conditions \cite{SotiriouFaraoni10}
\be\label{dSconditions}
f_0 ' R_0 =2f_0 \,. \;\;\;\;\;\;\; H_0=\sqrt{ \frac{f_0}{6f_0'}} 
\,.
\ee
The weak-field line element is written as
\begin{eqnarray}
ds^2 & = & -\left[ 1+2\Psi(r) -H_0^2r^2\right]dt^2 \nonumber\\
&&\nonumber\\
&+& \left[ 1+2\Phi(r) +H_0^2r^2 \right] dr^2 +r^2d\Omega^2 \,,
\end{eqnarray}
where $d\Omega^2=d\theta^2+\sin^2\theta d\varphi^2$ is the line 
element on the unit 2-sphere and $\Psi(r)$ and $\Phi(r)$ are 
post-Newtonian potentials. The goal is to compute these 
potentials by solving the linearized fourth order field equations
and to obtain the PPN parameter $\gamma=- \Psi/\Phi $, which is 
subject to the Cassini bound \cite{Bertottietal03}
\be
\left| \gamma -1 \right| <2.3 \cdot 10^{-5} 
\ee
(in GR, $\Psi=-\Phi$ is the Newtonian potential $-\kappa\, 
M/(8\pi r)$ and  
$\gamma=1 $).  A linearized 
analysis assuming that $ \left| \Psi(r) \right|, \left| \Phi(r) 
\right| \ll 1$, $H_0r \ll 1$, $f(R)$ is analytical at $R_0$, and 
$ mr \ll1$ yields \cite{Chibaetal, Olmo07,  
FaraoniTremblaycomment} 
\be \label{PPNpotentials}
\Psi(r)= -\frac{\kappa \, M}{6\pi f_0' r} \,, \;\;\;\;\;
\Phi(r)= \frac{\kappa \, M}{12\pi f_0' r} \,, \;\;\;\;\;
\gamma=\frac{1}{2} \,,
\ee
in gross violation of the Cassini bound. This result  would spell  
the end for  $f(R)$ gravity if it wasn't for the fact that the 
assumption of a light scalar field, $mr \ll 1$,  is 
violated.  In many $f(R)$ theories this happens naturally and the 
mass of 
$\phi$ is large at high ({\em i.e.}, Solar System) densities and 
almost zero  at cosmological densities  \cite{Faulkner}. But does 
this mechanism work for 
$f(R)=\alpha R^n$? 

To answer this question, note that for $n=2$ (the largest value 
of $n$ allowed by theoretical stability) the 
mass~(\ref{masssquared}) of the scalar 
$\phi=f'(R)$ in a de Sitter background is {\em exactly} zero and 
this field has infinite 
range independent of the density of the environment; 
therefore, it is certainly subject to Solar System 
constraints and $R^2$ gravity is ruled out experimentally.

At a first sight, it looks surprising that the mass $m$ vanishes 
while the potential~(\ref{potential}) turns out to be $V(\phi)= 
\frac{\phi^2}{4\alpha}$ for this theory. The solution to this 
apparent contradiction is that it is not $V(\phi)$, but rather 
the combination $\phi \frac{dV}{d\phi}-2V(\phi)$ that enters the 
equation of motion for the Brans-Dicke scalar \cite{BransDicke}
\be
\Box \phi=\frac{1}{2\omega+3} \left[ 8\pi T^{(m)}+\phi\, 
\frac{dV}{d\phi} -2V \right] \,,
\ee
where $T^{(m)}$ is the trace of the matter stress-energy tensor 
$T^{(m)}_{ab}$ (which, in the weak-field, slow-motion limit, 
reduces to  $-\rho$) and $\phi \frac{dV}{d\phi}-2V(\phi)$ 
vanishes 
identically for a purely quadratic potential \footnote{Because of 
this fact, the coefficient of $\phi^2$ in a quadratic potential 
should not be regarded as the mass of $\phi$ \cite{stambig}.}.

Incidentally, the theory $f(R)=\alpha R^2$ with $\alpha>0$ (in 
$D$ spacetime dimensions, $f(R)=\alpha R^{D/2}$ 
\cite{Vollick06})  has other peculiarities or theoretical 
problems  \cite{Ferrarisetal88, Sotiriou}:  it does not have the  
correct Newtonian limit \cite{PechlanerSexl66} and 
eq.~(\ref{dSconditions}) is satisfied for all, not for special, 
values of  the Ricci curvature $R$, which leads to unpleasant 
consequences \cite{Peter}. That something goes wrong in the 
weak-field limit can be seen in the post-Newtonian 
potentials~(\ref{PPNpotentials}) which, using $R_0=12H_0^2$, 
reduce to
\be
\Psi=-2\Phi=- \frac{ \kappa MH_0^{-1}}{864\pi \alpha} \, 
\frac{1}{H_0r}= - \frac{1}{216} \, \frac{ R_s cH_0^{-1}}{ 
\alpha} \, \frac{c H_0^{-1} }{r}
\ee
(restoring $G$ and $c$) where $R_s=2GM/c^2$ is the Schwarzschild 
radius of the mass $M$.  $\Psi$ and $\Phi$ are no longer 
guaranteed to be 
small in absolute value because  $  cH_0^{-1}/ r 
\ll 1$ and it is not clear how to choose the parameter  $\alpha$. 
A more refined analysis including terms of order $H_0r$ yields 
post-Newtonian potentials with Yukawa terms \cite{Faulkner, 
CapozzielloStabileCardone, Lubinietal}
\begin{eqnarray}
\Psi & = & -\frac{GM}{r}\left( 1-\frac{\delta \, \mbox{e}^{-ar} 
}{a^2r} \right) \,,\\
&&\nonumber\\
\Phi & = & \frac{GM}{r}\left( 1+\frac{\delta \, (1+ar) 
\, \mbox{e}^{-ar}}{a^2r} \right] \,.
\end{eqnarray}
In the limit $n\rightarrow 2^{-}$ in which $a\rightarrow 0$  
and  the range of the scalar 
becomes infinite, the Yukawa terms dominate the Newtonian ones 
and diverge. The range of $\phi$ {\em must} be kept small in 
order to  recover even the Newtonian limit \footnote{Note that 
these  considerations do not apply to $f(R)=R+\alpha R^n$ 
theories.}.

\begin{figure}
\scalebox{0.4}{\includegraphics{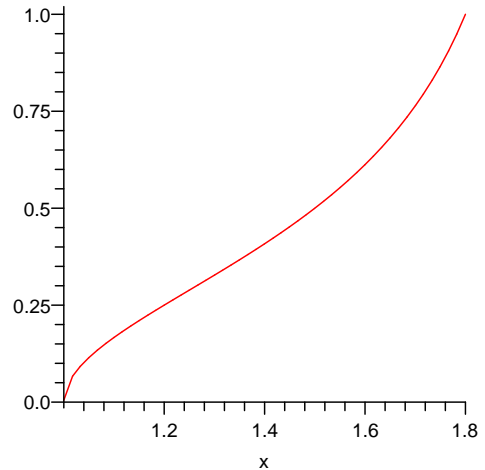}}%
\caption{\label{figA}The range $s(n)$ of the effective scalar 
degree of freedom $\phi=f'(R)$ (in units $cH_0^{-1}$) for the 
theory $f(R)=\alpha R^n$. The parameter $n$ is in the  
range  $1\leq n \leq  2$ allowed by stability. The function 
$s(n)$ starts out with vertical tangent at $n=1$.
}\end{figure}

At the opposite range of values for $n$ we have GR, which is 
viable and in agreement with all available Solar System 
experiments. Between the values $n=1$ and $n=2$, the range of the 
scalar field varies continuously but rapidly between zero and 
infinity. This range is given by the function 
\be\label{range}
 s(n) = \frac{1}{2}  \sqrt{\frac{n-1}{2-n}} \, c H_0^{-1} 
\ee
in the interval $\left[ 1,2 \right]$. This function varies 
continuously between 
$s(1)=0$ and its limit $\lim_{s\rightarrow 2^{-}} 
s(n)=+\infty$, always increasing. The derivative $s'(n)=  
\frac{cH_0^{-1}}{2\left( n-1 \right)^{1/2} \left( 2-n 
\right)^{3/2} } $ is always positive and the tangent to the graph 
of $s(n)$ starts vertically at $n=1$ and ends vertically as 
$n\rightarrow 2^{-}$, which means that the range of $\phi$ 
increases quickly as the $R^n$-theory departs very slightly from 
GR (see fig.~\ref{figA}). 
Clearly, as long as the exponent $n$ is very close to unity, the 
theory behaves as GR and passes  the experimental tests while, 
approaching values of $n$ closer to~2, the experimental bounds begin 
being violated, and disaster happens in the limit $n \rightarrow 
2^{-}$.  In conclusion, only for very small values of $n$ it is 
possible to invoke the  chameleon mechanism in  the 
weak-field analysis. By imposing the range of the scalar to be less 
than one astronomical unit ($1.496 \cdot 10^{13}$~cm) and using 
the value $H_0=70 \, \mbox{km}\cdot \mbox{s}^{-1}\cdot 
\mbox{Mpc}^{-1}$ for the Hubble parameter, 
one would obtain the requirement 
\be
0\leq \delta \equiv n-1 \leq 
5\cdot 10^{-30} \,.
\ee
Of course, realistic Solar System experiments 
do not have this level of precision, and the 
limit~(\ref{limitsondelta}) applies instead. This renders $R^n$ 
gravity a poor candidate for a realistic alternative to dark energy.

% figures should be put into the text as floats.
% Use the graphics or graphicx packages (distributed with LaTeX2e)
% and the \includegraphics macro defined in those packages.
% See the LaTeX Graphics Companion by Michel Goosens, Sebastian Rahtz,
% and Frank Mittelbach for instance.
%
% Here is an example of the general form of a figure:
% Fill in the caption in the braces of the \caption{} command. Put the label
% that you will use with \ref{} command in the braces of the \label{} command.
% Use the figure* environment if the figure should span across the
% entire page. There is no need to do explicit centering.

% Surround figure environment with turnpage environment for landscape
% figure
% \begin{turnpage}
% \begin{figure}
% \includegraphics{}%
% \caption{\label{}}
% \end{figure}
% \end{turnpage}

% tables should appear as floats within the text
%
% Here is an example of the general form of a table:
% Fill in the caption in the braces of the \caption{} command. Put the label
% that you will use with \ref{} command in the braces of the \label{} command.
% Insert the column specifiers (l, r, c, d, etc.) in the empty braces of the
% \begin{tabular}{} command.
% The ruledtabular enviroment adds doubled rules to table and sets a
% reasonable default table settings.
% Use the table* environment to get a full-width table in two-column
% Add \usepackage{longtable} and the longtable (or longtable*}
% environment for nicely formatted long tables. Or use the the [H]
% placement option to break a long table (with less control than 
% in longtable).
% \begin{table}%[H] add [H] placement to break table across pages
% \caption{\label{}}
% \begin{ruledtabular}
% \begin{tabular}{}
% Lines of table here ending with \\
% \end{tabular}
% \end{ruledtabular}
% \end{table}

% Surround table environment with turnpage environment for landscape
% table
% \begin{turnpage}
% \begin{table}
% \caption{\label{}}
% \begin{ruledtabular}
% \begin{tabular}{}
% \end{tabular}
% \end{ruledtabular}
% \end{table}
% \end{turnpage}

% Specify following sections are appendices. Use \appendix* if there
% only one appendix.
%\appendix
%\section{}

% If you have acknowledgments, this puts in the proper section head.
\begin{acknowledgments}
We thank Salvatore Capozziello for a discussion and the 
Natural Sciences and Engineering Research Council of Canada 
(NSERC) for financial support.
\end{acknowledgments}

% Create the reference section using BibTeX:
%\bibliography{simplified}

\end{document}